\documentclass{article}  
\usepackage{a4wide}
\pdfoutput=1
\usepackage[sectionbib,authoryear]{natbib}
\usepackage[pdftex]{graphicx}

\newtheorem{theorem}{Theorem}[section]
\newtheorem{lemma}[theorem]{Lemma}
\newtheorem{definition}{Definition}[section]
\newtheorem{corollary}{Corollary}[section]
\newenvironment{proof}[1][Proof]{\begin{trivlist}
\item[\hskip \labelsep {\bfseries #1}]}{\end{trivlist}}
\newcommand{\qed}{\nobreak \ifvmode \relax \else
      \ifdim\lastskip<1.5em \hskip-\lastskip
      \hskip1.5em plus0em minus0.5em \fi \nobreak
      \vrule height0.75em width0.5em depth0.25em\fi}

\usepackage{amsmath,amssymb}
\usepackage{color}
\usepackage{subfigure}

\begin{document}
\title{Sequence Annotation with HMMs:\\ New Problems and Their Complexity}
\def\institute{Faculty of Mathematics, Physics, and Informatics, 
           Comenius University,\\
           Mlynsk\'a Dolina, 842 48 Bratislava, Slovakia
}
      
\author{Michal N\'an\'asi
\and Tom\'a\v{s} Vina\v{r}
\and Bro\v{n}a Brejov\'a\\\institute}
\maketitle

\begin{abstract}
Hidden Markov models (HMMs) and their variants were successfully used
for several sequence annotation tasks. Traditionally, inference with HMMs is
done using the Viterbi and posterior decoding algorithms. However,
recently a variety of different optimization criteria and associated
computational problems were proposed. In this paper, we consider three
HMM decoding criteria and prove their NP hardness. These criteria
consider the set of states used to generate a certain sequence, but
abstract from the exact locations of regions emitted by individual
states. We also illustrate experimentally that these criteria are
useful for HIV recombination detection.

\end{abstract}

\section{Introduction}
Hidden Markov models (HMMs) and their variants were successfully used
for several sequence annotation problems in bioinformatics, including
gene finding, protein secondary structure prediction, protein family
modeling, detection of conserved elements in multiple alignments and
others \citep{Burge1997,Krogh2001,Siepel2005,Sonnhammer1997}.  
In many of these areas,
we assume that a particular biological sequence $X$ was generated by
the HMM, and we wish to infer which states of the model were used to
generate particular parts of the sequence in a process called HMM
decoding. The traditional algorithm for this task is the Viterbi
algorithm \citep{Viterbi1967}, which finds the state path (sequence of
states) generating sequence $X$ with the highest probability.

Many other decoding criteria were proposed \citep{Hamada2012}. 
For example, we
can assign labels to states of the HMM, and then search for the most
probable sequence of labels instead of the most probable state
path. If multiple states can share the same label, this problem is
NP-hard \citep{Lyngsoe2002,Brejova2007} 
and heuristics are used in practice \citep{Schwartz1990,Krogh1997}. 
In effect, we use state labels to group together many state paths 
with the same meaning and
then search for the group with the highest probability.  In some
application domains, it may be appropriate to group state paths
together in different ways.  In this paper, we explore three
optimization problems of this kind.

\begin{definition}[The most probable footprint]
The \emph{footprint} of a state path (or labeling) is the list of
states (or labels) visited on the path, discarding the information
about the number of successive characters emitted by the same state
(or label). The \emph{probability of a footprint} is the sum of
probabilities of all paths following the footprint. The task is to
find the most probable footprint for a given HMM and sequence.
\end{definition}

\begin{definition}[The most probable set]
The \emph{set} of a state path (or labeling) is the set of states (or
labels) visited on the path, regardless of their order or
multiplicity.  The \emph{probability of a set} is the sum of
probabilities of all paths sharing the same set. The task is to find
the set with the highest probability for a given HMM and sequence.
\end{definition}

\begin{definition}[The most probable restriction]
A path obeys a \emph{restriction} (set of states or labels) if it uses
only states or labels included in the restriction. The
\emph{probability of a restriction} is the sum of probabilities of all
paths that obey the restriction. The task is to find the restriction of
size $k$ with the highest probability for a given HMM and sequence.
\end{definition}

These problems were motivated by the HIV recombination detection
problem, which we review in Section \ref{sec:exp}. However, their use
is not limited to this application and is appropriate wherever
exact location of individual
regions in the sequence is not important. 
We demonstrate usefulness of these problems in practice even if
we use heuristics to solve them. Indeed, exact solution is
unlikely, since in
Sections \ref{sec:footprint}, \ref{sec:set} and \ref{sec:restriction},
we show that all three problems are NP-hard. 
The most probable footprint problem was briefly considered by
\citet{Brown2010}, who observe that it is polynomially solvable in HMMs
with two states or two labels. The other two problems were not studied
previously.

\paragraph{Hidden Markov models and notation.} 

In the rest of this section, we introduce the necessary notation. 
A \emph{hidden Markov model (HMM)} is a generative probabilistic model
with a finite set of states $V$ and transitions $E$. The generative
process starts by choosing a starting state $v_1$ according to the
initial state probabilities $I(v_1)$.  Then in each round, the model
emits a single symbol $x_i$ from the \emph{emission probability}
distribution $e(v_i,x_i)$ of the current state $v_i$, and then changes
the state to $v_{i+1}$ according to the \emph{transition probability}
distribution $a(v_i,v_{i+1})$. The process continues for
some fixed number of steps $n$.  Thus, the joint probability of
generating a sequence $X=x_1,\dots,x_n$ by a state path $\pi =
v_1,\dots,v_n$ in an HMM $H$ is $\Pr(\pi,X\mid H, n) = I(v_1) \cdot
e(v_1,x_1)\cdot \prod_{i=2}^n a(v_{i-1},v_{i}) \cdot e(v_i,x_i)$. In
other words, the HMM defines a probability distribution $\Pr(\pi,X
\mid H, n)$ over all possible sequences $X$ and state paths $\pi$ of
length $n$.

Let $Y$ be a sequence over some alphabet such that
$Y=x_1^{k_1}x_2^{k_2}\dots x_n^{k_n}$ where $x_{i}$ and $x_{i+1}$ are
distinct characters and each $k_j$ is greater than zero. 
Then the footprint $f(Y)$ of this sequence is $x_1x_2\dots x_n$ and
its character set $s(Y)$ is $\{x_1,x_2,\dots x_n\}$ (note that the size of this
set can be less than $n$). For example for $Y=aabaaacc$, we have $f(Y)=abac$
and $s(Y)=\{a,b,c\}$.

In particular, we will apply the footprint and set operators to
state paths $\pi$. Probability of a footprint $F$ for a given HMM $H$ and
sequence $X$ of length $n$ is 
$$\Pr(f(\pi) = F,X\mid H,n) = \sum_{\pi,f(\pi)=F} \Pr(\pi,X\mid H,
n).$$ Analogously we also define a probability of a given set of
states $S$ denoted as $\Pr(s(\pi) = S,X\mid H,n)$. Note that each path
$\pi$ included in this probability must use every state in $S$ at
least once. Finally, we will also discuss the probability of a state
restriction $S$ denoted as $\Pr(s(\pi) \subseteq S,X\mid
H,n)$, where we count all state paths that use only states from set
$S$, but are not required to use all of them.

We can also assign label $\ell(v)$ to each state $v$ of the HMM.  The
label $\ell(\pi)$ of a state path $\pi$ is then concatenation of
labels for individual states on the path. We can then use similar
notation for probability of footprints and sets defined on labelings,
such as $\Pr(f(\ell(\pi))=F,X\mid H,n)$.

We will say that a state path $\pi$ can generate $X$ if
$\Pr(\pi,X|H,n)>0$. Similarly a footprint $F$ can generate $X$ if
$\Pr(f(\pi) = F,X\mid H,n)>0$ and a set of states $S$ can generate $X$
if $\Pr(s(\pi) = S,X\mid H,n)>0$.

\section{Motivation}

\label{sec:exp}

The problems studied in this paper were inspired by the HIV
recombination detection problem, which was recently successfully
approached with jumping HMMs \citep{Schultz2006}. In this setting, we represent
sequence of each subtype of the HIV virus as a profile HMM, and
then we combine these profiles to a single HMM by addition of special
transitions modeling recombination between genomes of different
strains of the virus.  Given a particular genome, we try to
establish which portions were generated by which profile. However, it
is virtually impossible to determine the exact position of the
recombination. Therefore we may wish to group together state paths
that differ in positions of individual recombination points only by a
small amount \citep{Nanasi2010,Brown2010,Truszkowski2011}.

In this scenario, each subtype corresponds to one label.  Set of a
labeling $s(\ell(\pi))$ corresponds to the set of subtypes present in
the query sequence $X$. If we are not interested in the location of
recombination points, this is the most natural measure to
optimize. However, we might be interested to also know the order
of subtypes along the sequence represented by the footprint
of a labeling $f(\ell(\pi))$. 

Additionally, we can use a multi-step decoding strategy, where
we first fix a set of labels or a footprint, and then refine it
to a full labeling by a secondary optimization criterion. This
approach was taken by \citet{Truszkowski2011}, mainly as a heuristic
for speeding up the search. Here we show that this two-step strategy
can be also useful for improving the prediction accuracy. In particular,
as a second step we use the highest expected reward decoding (HERD) 
\citep{Nanasi2010}. The method  has two important
parameters: window size $W$ (breakpoints within this
distance are considered equivalent) and penalty $\gamma$ for
false positives (each true positive breakpoint is scored $+1$, false
positive breakpoint scores $-\gamma$). HERD optimizes
expected value of this scoring function under the assumption that the
sequence was generated from the HMM. 

\begin{figure*}[t]
\centerline{\includegraphics[width=0.8\textwidth]{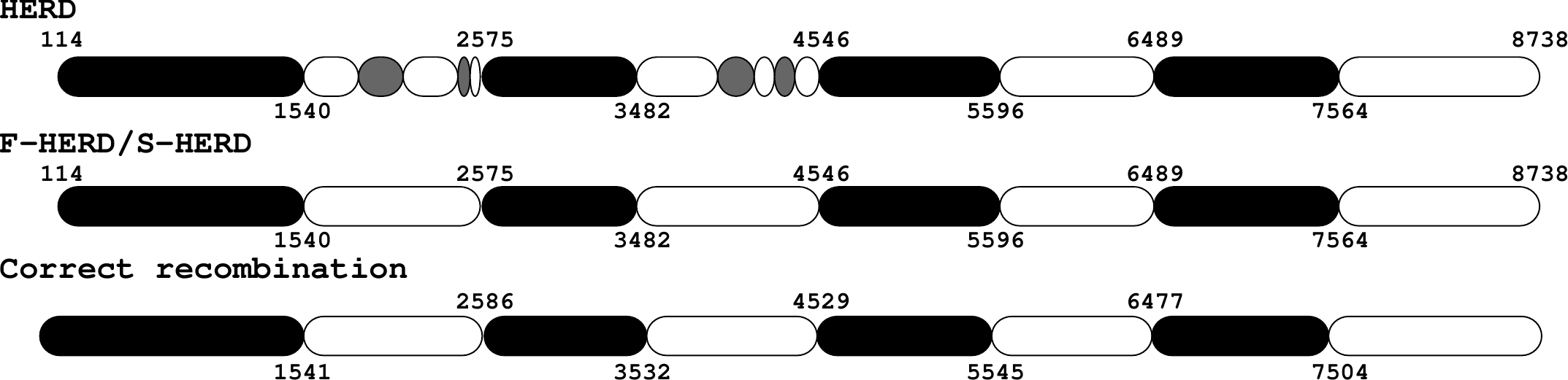}}
\caption{\label{fig:happy}
  Prediction of recombination on artificial recombinant of
  subtypes $A$ and $B$ (black and white) with recombination every
  950-1050 bases. HERD decoding yielded regions associated with incorrect
  subtypes (gray color representing 3 different subtypes) and
  fixing either the set or the footprint improved accuracy.}
\end{figure*}

\begin{figure*}[t]
\subfigure[]{
\includegraphics[width=0.43\textwidth]{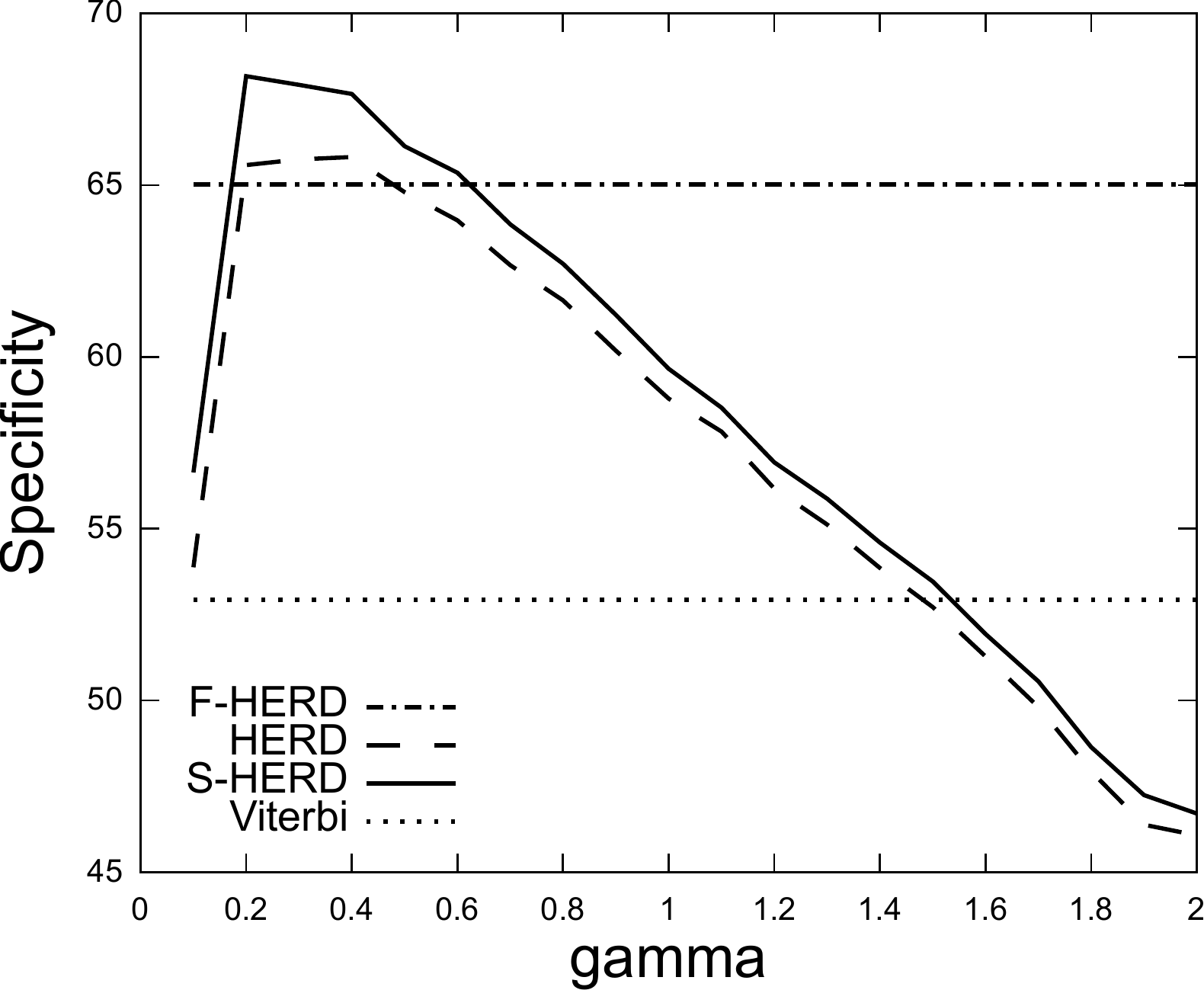}
\label{fig:subfig1}
}
\hfill
\subfigure[]{
\includegraphics[width=0.43\textwidth]{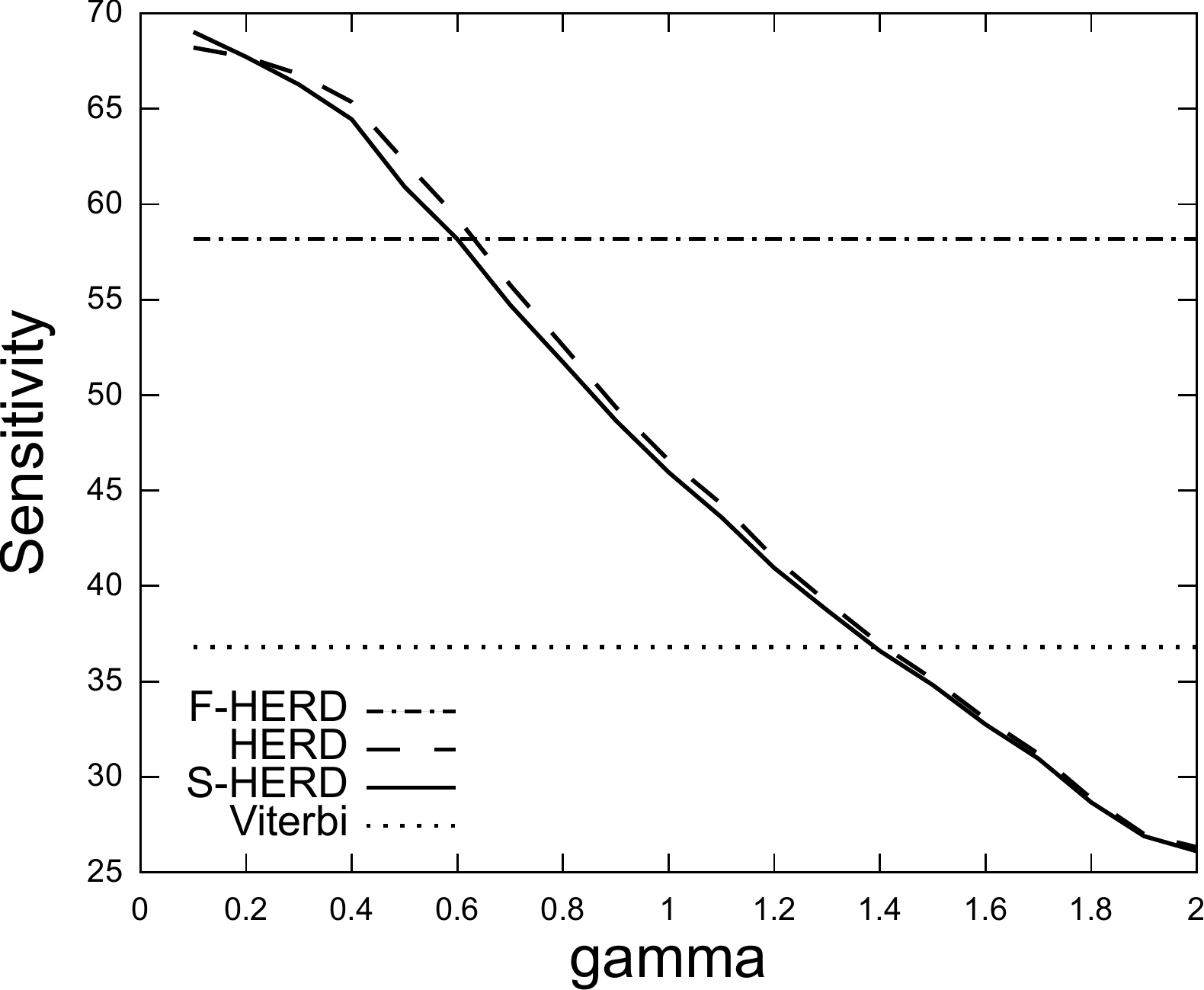}
\label{fig:subfig2}
}
\label{fig:subfigureExample}
\vglue-\baselineskip
\caption{\label{fig:experiment} Feature specificity (a) and 
  sensitivity (b) as a function of parameter $\gamma\in[0.1,2]$ on
  a semi-artificial set. A feature is correctly predicted if its boundaries
  are within 30 symbols of the corresponding feature in the correct
  annotation. Sensitivity is the proportion of real features that were
  correctly predicted and specificity is the proportion of predicted
  features that are correct. We use HERD parameters $P_j=10^{-5}$ and $W=10$.  
  F-HERD optimizes the same criterion among labelings obeying the footprint 
  obtained by sampling several paths
  from the probability distribution $\Pr(\pi\mid X,H,n)$, computing
  the footprint for each path, and then taking the most frequently
  occurring footprint among the samples, using the software by
  \citet{Truszkowski2011}. S-HERD optimizes HERD criterion among
  labelings using only labels from this footprint.
  The data set consists of 150 artificial recombinants of members of
  various subtypes of HIV virus with recombination every 200-300
  residues. 
}
\end{figure*}

As we can see in Figure
\ref{fig:experiment}, the program is very sensitive to the choice of
$\gamma$: for the optimal value of $\gamma$ it is significantly more
accurate than the Viterbi algorithm, but if we increase $\gamma$ too
much, the performance deteriorates. The most common problem is that
HERD predicts too many breakpoints when $\gamma$ is low 
(Figure \ref{fig:happy}).
By fixing a footprint as a constraint
in the two-step strategy, and then optimizing
the HERD criterion only for labelings obeying this footprint, the
prediction accuracy is virtually independent of $\gamma$ and
relatively close to the optimum values. Fixing the set instead of
the footprint yields slightly higher specificity and lower sensitivity
compared to optimizing HERD directly.  Note that 
the footprints and sets are chosen by a simple heuristic; 
perhaps even better results could be obtained 
with optimal choice of these constraints.

\section{The Most Probable Footprint}

\label{sec:footprint}

As previously seen, finding the most probable
footprint is a reasonable decoding criterion,
and it may also serve as a starting point in a multi-stage
strategy. In this section we show that this problem is
NP-hard. In particular, we will consider the footprint of a state path
$f(\pi)$. The problem of optimizing the footprint of a labeling
$f(\ell(\pi))$ is also NP-hard, because optimizing $f(\pi)$ is its special case, 
equivalent to optimizing $f(\ell(\pi))$ in an HMM in
which each state has a unique label.

\begin{theorem}
There is a fixed HMM $H$ such that the following problem is NP-complete:
Given a sequence $X$ of length $n$ and probability $p\in [0,1]$, determine
if there is a footprint $F$ such that $\Pr(f(\pi)=F,X\mid H,n)\ge p$.
\label{thm:foot}
\end{theorem}

\begin{proof}
We will prove NP-hardness by a reduction from the maximum clique problem
using the HMM in Figure 
\ref{fig:footprint_hmm} with eight states and
alphabet $\Sigma=\{S,S',T,T',\#,0,1,?\}$. 

\begin{figure*}[t]
\centerline{\includegraphics[scale=0.68]{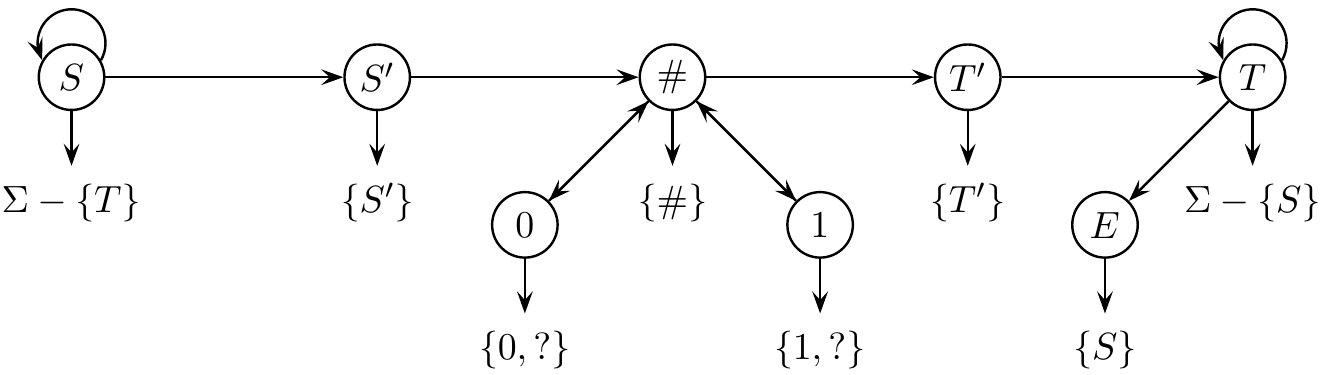}}
\caption{\label{fig:footprint_hmm}
The HMM from the proof of Theorem \ref{thm:foot}. Each
circle denotes one state. The HMM always starts in state $S$.  
Under each state is the set of symbols that the state emits with non-zero
probability. Each of these symbols is emitted with probability $1/k$, 
where $k$ is the size of the set. 
Alternatively, all outgoing transitions from a particular state have the
same probability.}
\end{figure*}

%Popis zakodovania grafu
Let $G=(V,E)$ be an undirected graph with $n$ vertices
$V=\{1,2,\dots,n\}$. We will encode it in a sequence $X$ over alphabet $\Sigma$ as follows.
For every vertex $v\in V$, we create a block $X_v$ with 
$2n+3$ symbols: $X_v=S'\#b_{v,1}\#b_{v,2}\#\dots\#b_{v,n}\#T'$ where $b_{i,j}=1$ 
if $i=j$, $b_{i,j}=?$ if $(i,j)\in E$ and $b_{i,j}=0$ otherwise. 
Sequence $X$ is a concatenation of blocks for all vertices 
with additional first and last symbols: $X=SX_1X_2\dots X_nT$.

All state paths that can generate $X$ have a similar structure. The first
symbol $S$ and several initial blocks are generated in state $S$, one
block, say $X_i$, is generated in states $S'$, $\#$, $0$, $1$, and
$T'$ and the rest of the sequence, including the final symbol $T$ is
generated in state $T$. We will say that a state path with this
structure \emph{covers} the block $X_i$.  Note that state $E$ is never used in
generating $X$, its role is to ensure that the probability of
self-transition is the same in states $S$ and $T$.
All state paths that can generate $X$ have the same
probability $q = \Pr(\pi,X\mid H,|X|) = 2^{-2n^2-2n}3^{-n-1}7^{-2n^2-n+1}$.

%celkovo 2n^2+3n+2 symbolov
%1/2 zacni v S
%kazdy symbol v S a T ma 1/7 emisiu a 1/2 tranziciu do dalsieho
% tych symbolov je 2n^2+n-1
%prechody z S' a T', 0, 1 su 1, tych je n+2
%prechody z # so 1/3, tych je n+1
%emisia v S' a T' a # je 1, tych je n+3
%emisia v 0 a 1 je 1/2, tych je n
%
%spolu teda 1/2^{1 + 2n^2+n-1 + n} * 1/3^{n+1} * 1/7^{2n^2+n-1} 

We say that a state path $\pi$ is a \emph{run} of footprint $F$, if
$\pi$ can generate $X$, and $f(\pi)=F$.
Every footprint $F$ that can generate $X$ has the following structure:
$F=SS'\#c_1\#c_2\#\dots\#c_n\#T'T$ where $c_i\in\{0,1\}$. 
The probability of footprint $F$ is $qk$ where $k$ is the number of its runs.
Also note that every run of $F$ covers a different $X_i$, because once $X_i$
is known, the whole path is uniquely determined. 

We will now prove that the graph $G$ has a clique of size at least $k$
if and only if there is a footprint for sequence $X$ with
probability at least $qk$.  First, let $R$ be a clique in $G$ of size
at least $k>0$.  Consider the footprint
$F=SS'\#c_1\#c_2\#\dots\#c_n\#T'T$ where $c_i=1$ if $i\in R$ and $c_i=0$
otherwise. For any $i\in R$, there is a run $\pi_i$ of $F$ that covers
$X_i$. This run will use state 1 for generating each $b_{i,j}$ such
that $j\in R$ and thus both $b_{i,j}\in \{?,1\}$ and $c_j=1$.  For
$j\notin R$ we have $b_{i,j}=0$ and $c_j=0$, thus they will use state
0 in $\pi$. Since there is a different run for every $i\in R$, footprint 
$F$ has at least $k$ runs.

Conversely, let $F$ be a footprint with probability at least $qk>0$
and thus with at least $k$ runs. We will construct a clique of size at
least $k$ as follows. Let $R$ be the set of all vertices $i$ such that
$f$ has a run that covers $X_i$. Clearly the size of $R$ is at least
$k$.  Since $F$ has non-zero probability, it has the form
$SS'\#c_1\#c_2\dots\#c_n\#T'T$ for $c_i\in \{0,1\}$. For all $i\in R$,
$c_i=1$ because the $i$-th block has $b_{i,i}=1$. Therefore for all
$i,j\in R$, we have $b_{i,j}\in \{1,?\}$, which means that $(i,j)\in
E$ or $i=j$. This implies that $R$ is indeed a clique.

To summarize, given graph $G$ and threshold $k$, we can compute in
polynomial time sequence $X$ and threshold $qk$ such that $G$ has a
clique of size at least $k$ if and only if sequence $X$ has a
footprint with probability at least $qk$. This completes our reduction.

The problem is in NP (even if HMM is not fixed, but given on input),
because given an HMM $H$, sequence $X$ and a footprint $F$, we can
compute the probability $\Pr(f(\pi)=F,X\mid H,|X|)$ in polynomial time
by a dynamic programming algorithm which considers all prefixes of
$X$ and all prefixes of $F$. If probability $p$ and parameters of HMMs
are given as rational numbers, we can compute all quantities without
rounding in polynomial number of bits. \qed
\end{proof}

\section{The Most Probable Set of States}
\label{sec:set}

In this section, we prove NP-hardness of finding the most probable set
of states. Again, as with footprint, this is a special case of the
problem of finding the most probable set of labels. 

\begin{theorem}The following decision problem is NP-hard:
Given an HMM $H$, sequence $X$ of length $n$, and a number $p\in [0,1]$,
decide if there exists a set of states $S$ such that 
$\Pr\left(s(\pi)=S, X\mid H, n \right)\geq p$.
\end{theorem}\label{thm:sets}

To prove this theorem, we will use a reduction from the maximum clique
problem.  Given a graph $G=(V,E)$ and a clique
size $k$, we first choose a suitable threshold $k'\ge k$, as
detailed below, and construct a graph $G'=(V',E')$ such that $G'$ has
a clique of size $k'$ if and only if $G$ has a clique of size
$k$. This is achieved simply by adding $k'-k$ new vertices and
connecting each of the new vertices to all other vertices in $V'$.
As long as $k'-k$ is not too large, this transformation can be done in
polynomial time.

In the next step, we use $G'$ and $k'$ to construct an HMM, an input
sequence and a probability threshold. We will use the following
straightforward way of converting a graph to an HMM.

\begin{definition}\label{GraphHMM}
Let $G=(V,E)$ be an undirected graph (without self-loops). 
Then the \emph{graph HMM} $H_G$ is defined as follows:
\begin{itemize}
\item Its set of states is $V\cup \{\psi\}$, where $\psi\notin V$ is a
  new state called the error state.
\item Its emission alphabet is $\{0,1\}$.
\item Each state $v\in V$ has initial probability $I(v) = 1/|V|$, the
error state has initial probability $I(\psi)=0$.
\item Each state $v\in V$ emits 0 with probability 1, the error state emits 1 
with probability 1.
\item Transitions with non-zero probability between states $u,v\in V$
  correspond to edges in $E$, more precisely:
$$a(u,v)=\begin{cases}
\frac1{|V|} & \{u,v\}\in E\\
0 & \text{otherwise}
\end{cases}$$
\item For $u\in V$, we also have $a(u,\psi)=1-\sum_{v\in V}a(u,v)$
and $a(\psi,u)=0$. The error state has a self-transition with 
probability 1: $a(\psi,\psi)=1$.
\end{itemize}
\end{definition}

The error state $\psi$ is added to the HMM so that all non-zero
transitions between states in $V$ have the same probability. Any state
path $\pi$ containing only states from $V$ connected by transitions
with non-zero probability has the same probability of generating
sequence $X=0^n$: $\Pr(\pi, X=0^n|H_G,n) = |V|^{-n}$.
Such paths correspond to walks in graph $G$. 

Therefore, we will be interested in counting the number of walks in
different graphs. Let $Y(n,G)$ be the number of walks of length $n-1$
in a graph $G=(V,E)$ that visit every vertex from $G$ at least once.
Note that a walk of length $n-1$ contains $n-1$ edges and $n$ vertices, and
therefore $Y(n,G)=0$ for $n<|V|$. As a special case we consider
$D(n,k)=Y(n,K_k)$, where $K_k$ is the complete graph with $k$
vertices. The following claim clearly holds:

\begin{lemma}\label{NotCliqueIsSmaller}
If $G$ is a graph with $k$ vertices and $n\ge k$, then
$Y(n,G)\le D(n,k)$ with equality only for $G=K_k$. 
\end{lemma}

In our reduction we use HMM $H=H_{G'}$ and $X=0^n$ for a suitable
choice of $n$ discussed below. As threshold $p$ we will use the value
$D(n,k')/|V|^{n}$. Clearly, if the input graph $G$ has a clique $S$ of
size $k$, graph $G'$ has a clique $S'$ of size $k'$. There
are at least $D(n,k')$ walks of length $n-1$ that use only vertices in
$S'$ and visit each of them at least once. Each of such walks
corresponds to one state path, and therefore the probability of
the set of states $S'$ is exactly $p$. 

In order to prove the opposite implication, we need suitable choices
of $n$ and $k'$. Table \ref{DNKTable} shows values of $D(n,k)$ for
small values of $n$ and $k$. For a fixed length of walk $n$, the
number of walks in $K_k$ initially grows with increasing $k$, as we
have more choices which vertex to use next, but as $k$ approaches $n$,
$D(n,k)$  may start to decrease, because the walks are more constrained by
the requirement to cover every vertex. We are particularly interested in
the value of $k$ where $D(n,k)$ achieves the maximum value for a fixed $n$. 
In particular we use the following notation:
$$M_{n} = \min\left\{k ; D(n,k) = \max_{0\leq k'\leq
  n}D(n,k')\right\}$$ Note that if there are multiple values
of $k$ achieving maximum, we take the smallest one as $M_n$.  In our
reduction, we would like to set $n$ to be the smallest value such that
$M_n=k$, but we were not able to prove that such $n$ exists for each $k$.
Therefore we choose as $n$ the smallest value such that $M_n\ge k$, and 
we denote this value $n_k$. As $k'$ we then use $M_{n_k}$. The following 
lemma states important properties of $n_k$ and $M_{n_k}$. 

\begin{table*}[t]
\begin{center}
\begin{tabular}{|c||c|c|c|c|c|c|c|c|c||c|}\hline
\bf n/k&\bf 0&\bf 1&\bf 2&\bf 3&\bf 4&\bf 5&\bf 6&\bf 7&\bf 8&\bf
$\mathbf{M_n}$\\\hline\hline
\bf 0&{\bf 1}&&&&&&&&&0\\\hline
\bf 1&&\bf 1&&&&&&&&1\\\hline
\bf 2&&&\bf 2&&&&&&&2\\\hline
\bf 3&&&2&\bf 6&&&&&&3\\\hline
\bf 4&&&2&18&\bf 24&&&&&4\\\hline
\bf 5&&&2&42&\bf 144&120&&&&4\\\hline
\bf 6&&&2&90&600&\bf 1200&720&&&5\\\hline
\bf 7&&&2&186&2160&7800&\bf 10800&5040&&6\\\hline
\bf 8&&&2&378&7224&42000&100800&\bf 105840&40320&7\\\hline
\bf 9&&&2&762&23184&204120&756000&\bf 1340640&1128960&7\\\hline
\bf 10&&&2&1530&72600&932400&5004720&13335840&\bf 18627840&8\\
\hline\hline
$\mathbf{n_k}$&0&1&2&3&4&6&7&8&10&\\\hline 
$\mathbf{M_{n_k}}$&0&1&2&3&4&5&6&7&8&\\\hline 
\end{tabular}
\end{center}
\caption{Values of $D(n,k)$, $n_k$, $M_n$, and $M_{n_k}$ for small values of $n$ and $k$. Empty cells contain zeros.}\label{DNKTable}
\end{table*}

\begin{lemma}\label{DNKlemma}
The value of $n_k$ is at most $\lceil k\ln k\rceil$ and $n_k$ and
$M_{n_k}$ can be computed in $O(k^{O(1)})$ time.
\end{lemma}

Before proving this lemma, we finish the proof of the reduction. Let
us assume that there is a set of states $S$ such that
$\Pr(s(\pi)=S,X|H,n)\ge p$. This means that if we consider walks in the
subgraph $G'(S)$ induced by the set $S$, we get $Y(n,G'(S))\ge
D(n,k')$. We will consider three cases:
\begin{itemize}
\item If $S$ is a clique and $|S|\ge k'$, we have the desired clique in 
graph $G'$, and therefore there is also a clique of size $k$ in graph $G$. 
\item If $S$ is a clique and $|S|<k'$, then by definition of $M_n$ we have 
$Y(n,G'(S))=D(n,|S|)<D(n,M_n) = 
D(n,k')$. This is a contradiction with our assumption. 
\item If $S$ is not a clique, then by Lemma \ref{NotCliqueIsSmaller}
  and definition of $M_n$ we have $Y(n,G'(S)) < D(n, K_{|S|}) \le
  D(n,M_n) = D(n,k')$. Again we get a contradiction with 
the inequality $Y(n,G'(S))\ge D(n,k')$.
\end{itemize}
Therefore we have proved that $G$ contains a clique of size $k$ if and
only if the most probable set of states in $H_{G'}$ that can generate $X$ has
probability at least $p$. Moreover, we can construct $n_k$, $M_{n_k}$,
$H_{G'}$, $X$, and $p$ in polynomial time.

To complete this proof we need to prove Lemma \ref{DNKlemma}.  We
start by proving another useful lemma.

\begin{lemma}\label{RecurenceLemma}
For $2\leq k\leq n$ the following recurrence holds:
$$D(n,k)=(k-1) D(n-1,k) + k D(n-1,k-1).$$
In addition, $D(n,n)=n!$, $D(n,1) = 0$ for $n>1$, and $D(n,k) = 0$ for $k>n$.
\end{lemma}
\begin{proof}
Clearly, $D(n,n)=n!$ since walks of length $n-1$ correspond to
permutations of vertices.  If $n>1$ then $D(n,1)=0$, since $K_1$ does
not contain any edges.  If $k>n$, $D(n,k)=0$ since a walk of length
$n-1$ can pass through at most $n$ vertices.

Now let $2\leq k\leq n$. Denote as $v(w)$ the number of different
vertices covered by walk $w$. Let $w$ be a walk of length $n-1$ with
$v(w)=k$ and let $w'$ be a walk obtained by taking the first 
$n-1$ vertices of walk $w$. Then $v(w')$ is either $k$ or $k-1$. 

Every walk $w'$ of length $n-2$ with $v(w')=k$ can be extended to a
walk $w$ of length $n-1$ in $K_k$ in $k-1$ ways, because as the last vertex
of $w$ we can use any vertex except the last vertex of $w'$. Therefore
there are $(k-1) D(n-1,k)$ different walks $w$ in $K_k$ with
property $v(w')=k$.

On the other hand if $v(w')=k-1$, we can create a walk $w''$ in
$K_{k-1}$ by renumbering the vertices in $w'$ so that only numbers
$\{1,\dots,k-1\}$ are used (if the vertex missing in $w'$ is $i$, we
replace $j$ by $j-1$ for every vertex $j>i$).  The same
representative $w''$ is shared by $k$ different walks $w$, because to
create $w$ from $w''$, we need to choose the missing vertex $i$ from
all $k$ possibilities, renumber vertices to get $w'$ and then to add
the missing vertex $i$ at the end of the walk. Therefore there are
$kD(n-1,k-1)$ walks with the property $v(w')=k-1$. Combining the two
cases we get the desired recurrence.\qed
\end{proof}

\begin{proof}[Proof of lemma \ref{DNKlemma}] 
Assume that $k\ge 3$. Clearly, $D(n,k)\leq k(k-1)^{n-1}$, since
$k(k-1)^{n-1}$ is the number of all walks of length $n-1$ in
$K_k$. However, this number includes also walks avoiding some
vertices. The number of such walks can be bounded from above by
$k(k-1)(k-2)^{n-1}$ where we choose one of the $k$ vertices to avoid
and then consider all possible walks on the remaining $k-1$
vertices. In this way we count some walks multiple times, nonetheless
by Bonferroni inequality we obtain bound $D(n,k)\geq
k(k-1)^{n-1}-k(k-1)(k-2)^{n-1}$.

For $k\ge 4$ we therefore have that if
$(k-1)(k-2)^{n-1}<k(k-1)^{n-1}-k(k-1)(k-2)^{n-1}$, then
$D(n,k-1)<D(n,k)$.  By taking logarithm of both sides of the inequality we
obtain $n>f(k)$ where $f(k) = 1+\frac{\ln(k^2-1)-\ln
  k}{\ln(k-1)-\ln(k-2)}$.  Let $n = \lceil f(k)\rceil$ for some $k\ge
4$ and consider row $n$ in Table \ref{DNKTable}.
We have that $D(n,k-1)<D(n,k)$ and since function $f$ is
increasing, we also we have that $D(n,k'-1)<D(n,k')$ for all $k'\le k$ 
(we have proved it only for $k'\ge 4$, but it is easy to see that it is
also true for $2\le k'\le 3$). The maximum in row $n$ is therefore
achieved at some position $M_n \ge k$. Recall, that $n_k$ is the
smallest $n$ such that $M_n\ge k$. Therefore $n_k\leq \lceil f(k)\rceil$.
The function $k \ln k/f(k)$ is decreasing and its limit is $1$ as $k$
approaches $\infty$. Therefore $\lceil f(k)\rceil\leq\lceil k\ln k\rceil$,
which gives us the inequality $n_k\le \lceil k \ln k\rceil$. This inequality 
can also be easily verified for $k<4$. Since $M_n\le n$,
we also have $M_{n_k}\le \lceil k \ln k\rceil$. 

We can compute $n_k$ and $M_{n_k}$ by filling in table $D(m,j)$ for
all values of $m$ and $j$ up to $\lceil k\ln k\rceil$ using the
recurrence from lemma \ref{RecurenceLemma}. Since $D(n,k)\leq k^n\leq
n^n$, we can store $D(m,j)$ in $O(k \mbox{polylog}(k))$ bits.
Therefore computing the desired values $n_k$ and $M_{n_k}$ 
can be done in polynomial time. \qed
\end{proof}

By using the same reduction as in Theorem \ref{thm:sets}, we can
also prove NP-hardness of the following variant of the problem, in
which we restrict the size of the set of states $S$. 

\begin{corollary}
The following problem is NP-hard:
Given is an HMM $H$, sequence
$X$ of length $n$, integer $k$ and a number $p\in [0,1]$ and the task to
decide if there exists a set of states $S$ of size exactly $k$ 
such that $\Pr\left(s(\pi)=S, X\mid H, n \right)\geq p$.
\end{corollary}

%We can easily see that the existence of a clique
%of size $k$ in $G$ leads to the existence of the desired set of states
%$S$ of size exactly $k'$, and by following the reasoning from the
%proof, we can also prove that the converse is true.

Note that it is not clear if the most probable set of states problem is in
NP. In particular, given a set of states $S$, it is NP-hard to find out if its
probability is greater than some threshold $p$, even if this threshold
is 0, as we show next.

\begin{theorem}
Given HMM $H$, sequence $X$ of length $n$ 
and a subset of state space $S$, the problem of deciding if
$\Pr\left(s(\pi)=S, X\mid H, n\right)$ is non-zero is NP-complete.
\end{theorem}

\begin{proof}
Let $G=(V,E)$ be a graph  
and $H_G$ be the corresponding graph HMM as
in Definition \ref{GraphHMM}. Let $X=0^{|V|}$.  Any state path
that can generate $X$ and contains all vertices from $V$ contains each
vertex exactly once.  It is easy to see that $\Pr\left(s(\pi)=V,X \mid H_G, |X|
\right)>0$ if and only if $G$ contains a Hamiltonian path. \qed
\end{proof}

Unlike the most probable footprint problem, which was NP-hard even for
a fixed HMM of a constant size, the most probable set problem is
fixed-parameter tractable with respect to the size of the HMM. Given
an HMM with $m$ states and a sequence of length $n$, we can find the
most probable set of states in time $O(2^m m^2 n)$ by a dynamic
programming algorithm similar to the Forward algorithm.
We define $F[i,S,v]$ to be the sum of probabilities of all
states paths $\pi$ of length $i$ such that $s(\pi)=S$, $\pi$ ends in
state $v$ and generates the first $i$ characters of sequence $X$.
To compute $F[n,S,v]$ we use the following equation:
\ifx\settwocolumn\undefined
$$F[i,S,v] = \begin{cases}
I(v)e(v,X[1])& i=1,S=\{v\}\\ 
\displaystyle \sum_{u\to v}a(u,v)e(v,X[i])\left(F[i-1,S\backslash\{v\},u]
+ F[i-1,S,u]\right) & i>1, v\in S\\
0 & \mbox{otherwise}
\end{cases}$$
\else
$$F[i,S,v] = \begin{cases}
I(v)e(v,X[1])& i=1,S=\{v\}\\ 
\displaystyle \sum_{u\to v}a(u,v)e(v,X[i])\\\hspace{5mm}\left(F[i-1,S\backslash\{v\},u]\right.
\\\hspace{5mm}\left.
+ F[i-1,S,u]\right) & i>1, v\in S\\
0 & \mbox{otherwise}
\end{cases}$$
\fi

\section{The Most Probable State Restriction}

\label{sec:restriction}

In the most probable set problem, we consider only paths that use each
state in the set. In some situations it is more natural to allow
paths to use only some of these states, as in the most probable
restriction problem. However, the full set of states of the model is
trivially the most probable restriction. To get a meaningful problem
definition, we restrict the size of the restriction to be $k$. As we
will show, this problem is also NP-hard.

\begin{theorem}
The following problem is NP-complete: Given is an HMM $H$, sequence
$X$, integer $k$ and number $p\in [0,1]$. Determine if
there is a subset of states $S$ of size $k$ such that 
$\Pr(s(\pi)\subseteq S, X\mid H,|X|)\ge p$. 
\end{theorem}

\begin{proof}
We will prove NP-hardness by a reduction from 3-SAT. Consider an
instance of 3-SAT with the set of variables $U=\{u_1,u_2,\dots,u_n\}$
and the set of clauses $C=\{c_1,c_2,\dots,c_m\}$. 
Based on sets $U$ and $C$, we construct an HMM $H$ as
follows.  The set of states $V$ will contain all positive and negative
literals. The emission alphabet $\Sigma$ contains all clauses, all
variables and a special error symbol $\psi$. The initial probability
$I(v)$ of each state is $1/(2n)$, and the transition probability
$a(u,v)$ between any two states is also $1/(2n)$. State for a literal
$u$ emits with probability $1/|\Sigma|$ every clause that contains $u$.
State for literal $u$ also generates the positive form of the literal
with probability $1/|\Sigma|$. Finally, to achieve the sum of emission
probabilities to be one, it also generates the error symbol 
with probability $1-\sum_{x\in C\cup U}e(v,x)$. 

Based on the SAT instance, we also create string $X=u_1u_2\dots
u_nc_1c_2\dots c_m$ and set the size of the restriction $k$ to equal
the number of variables $n$. Every state path $\pi$ that can generate
$X$ has probability $(2n|\Sigma|)^{-|X|}$, we set threshold $p$ to
this value. The first part of sequence $X$ contains all variables, and
variable $u_i$ can be generated only by states $u_i$ and
$\bar{u_i}$. Therefore one of these two states needs to be in the
path. Since the first portion of the path already traverses $k$ 
different states, only these states can be used to emit the second part of the
sequence. Every clause can be emitted only by states for literals that
satisfy it. The set of states used by a particular state path with
non-zero probability therefore corresponds to a satisfying assignment
in a straightforward way. The HMM has a restriction of size $k$ with
probability at least $p$ if and only if the 3-SAT instance has a
satisfying assignment.

Note that given a restriction $S$, we can easily verify if its
probability is at least $p$ by a variant of the Forward algorithm
in which we allow only states in $S$. Therefore the problem is in NP.
\qed
\end{proof}

\section{Conclusion}

In this paper, we have proved NP-hardness of three HMM decoding
problems. 
The most probable footprint problem can be viewed as a special case of
the most probable ball problem under the border shift distance 
considered by \citet{Brown2010}. In
this problem, we sum probabilities of all labelings that have the same
footprint and differ in positions of all feature boundaries by at
most $d$.
\citet{Brown2010} observe that if the HMM is allowed to contain
multiple states of the same label, the most probable ball problem is
NP-hard even for $d=0$. If $d\ge n$, where $n$ is the length of the
input sequence, the most probable ball problem is equivalent to the
most probable footprint problem. Therefore, our results imply NP
hardness of the most probable ball problem for large values of $d$
even in HMMs in which each state has a unique label. However, it is
open if the problem is NP hard even for small values of $d$ in such HMMs.

In spite of their hardness, we have demonstrated that the studied problems 
do have practical applications, even if we have to resort to heuristics in order to solve them.
From a practical point of view, it would be useful to explore better heuristic
approaches, or even approximation algorithms with
provable bounds. It is also of interest to study if polynomial
algorithms exist for some special classes of HMMs. For example, as
pointed out by \citet{Brown2010}, the most probable footprint problem
is polynomially solvable in HMMs with two states or two labels, because
a sequence of length $n$ has only $2n$ possible footprints.  

%Finally,
%we have proved that the most probable set problem is NP-hard, but it
%is open, if this problem actually belongs to NP.

\paragraph{Acknowledgments.} This research was supported 
by European Community FP7 grants IRG-224885 and IRG-231025, grant 
1/1085/12 from VEGA and Comenius University grant UK/465/2012.

%\paragraph{Acknowledgements.} This research was partially supported 
%by European Community FP7 grants IRG-224885 and IRG-231025, 
%grant VEGA xxx and Comenius University grant yyy.

\bibliographystyle{apalike} \bibliography{main}

\end{document}